\documentclass[aps,pra,showpacs,amssymb,preprint, nobibnotes,]{revtex4-1}
\newcommand{\beq}{\begin{equation}}
\newcommand{\eeq}{\end{equation}}
\usepackage[dvips]{graphicx,graphics}
\usepackage{epsfig}

\begin{document}

\title{Fast and efficient transport of  large ion clouds}

\author{M.R. Kamsap}
\author{ J. Pedregosa-Gutierrez}
\author{ C. Champenois}
\email{caroline.champenois@univ-amu.fr}
\author{ D. Guyomarc'h}
\author{ M. Houssin}
\author{ M. Knoop}

\affiliation{Aix-Marseille Universit\'e, CNRS, PIIM, UMR 7345, 13397 Marseille, France}

\date{\today}

\begin{abstract}
The manipulation of trapped charged particles by electric fields is an accurate, robust and reliable technique for many applications or experiments in high-precision spectroscopy. The transfer of the ion sample between multiple traps allows the use of a tailored environment in quantum information, cold chemistry, or frequency metrology experiments. In this article, we experimentally study   the transport of  ion clouds  of up to 50 000 ions. The design of the trap makes  ions very sensitive to any mismatch between the assumed  electric potential and  the actual local one. Nevertheless, we show that being fast (100~$\mu$s to transfer over more than 20~mm) increases the  transport efficiency to values higher than 90~\%, even with a  large number of ions. For clouds of less than 2000 ions, a 100~\% transfer efficiency is observed. \end{abstract}

\maketitle

\section{Introduction}\label{s_intro}

An ion cloud confined in a linear radiofrequency (RF) quadrupole trap is an example of a non-neutral plasma (NNP),  a plasma consisting exclusively of particles with a single sign of charge. Its thermal equilibrium state has been studied in detail by Dubin, O'Neil and coworkers in the context of large ensembles in Penning traps (for a complete review, see \cite{dubin99}) and extrapolated to ions in RF-quadrupole traps \cite{dubin93, schiffer00}  and even to ions in multipole traps \cite{champenois09}. Thanks to Doppler laser cooling, the transition of an ion cloud to a liquid and a Coulomb crystal phase was observed \cite{bluemel88,drewsen98} after it was theoretically studied \cite{dubin93,schiffer93}.

In this paper, we address an out-of-equilibrium issue with the transport of ion clouds by the translation and deformation of the trapping potential. Our aim is to shuttle ions between separate trapping zones, and the objective is  to do this as fast as possible without loss.  This problem is related to the transport of single ions in micro-traps, which has been realized mainly for quantum information processing (QIP) applications. In our experiment, a more general problem is studied as the transported ensemble is a many-body system with long-range interaction. However,  we can compare the center-of-mass motion of the ion cloud to the trajectory of a single ion, and we therefore use the QIP transports as model systems. 

 Different groups have addressed the question of transport of single ions \cite{reichle06, schulz06, hucul08} as it is a crucial issue for scalable architectures of QIP  in ion traps.  One of the main concerns in these experiments is to avoid heating issues during the transport, the implementation of robust gate operations also requires very high transfer efficiencies \cite{wright13}. In micro-traps, the transport distances for a single ion are of the order of 100$\mu$m. Care is taken to translate the ion in a quasi-constant potential well, which requires a large number of electrodes to tailor the trapping potential along the transport path. Speed is an additional issue which has to be taken into account, as shuttling ions between different sites is only a preparatory or intermediate task in a more sophisticated protocol and should not last longer than the computational gate. 

Our experiment comprises two  macroscopic linear quadrupole RF traps, storing ion clouds  of  a thousand  up to a million ions. The  trapping and shuttling along the common trap $z$-axis are controlled by three DC-electrodes.   Such large traps are typically used in frequency metrology in the microwave domain \cite{prestage07}, exotic ion studies \cite{herfurth03} or experiments in physical chemistry \cite{wester09}. Two different trapping zones are useful to keep one zone free from contact potential induced by neutral atom deposit or to accumulate ions in one of the trapping zones. High efficiencies for transfer are mandatory, and rapid transport protocols can reduce dead times which are detrimental to frequency stability in the case of atomic clocks and to sample conservation in case of short-lived species.

Many-body transport is also a concern for experiments with an  ensemble of cold neutral atoms  and Bose-Einstein condensates (BECs) which have been transported without heating, making use of shortcuts to adiabacity \cite{couvert08,schaff11b}. In \cite{couvert08} a cloud of a few $10^6$ cold atoms is shuttled back and forth with an optical tweezer over a distance of 22.5 mm, in times as short as four trap oscillation periods. The use of an  optical tweezer is very advantageous as it can be moved without deformation and the faster than adiabatic transport scheme relies on this non-deformation. The scheme designed in \cite{schaff11b} allowed the authors to translate a cold gas in the non-interacting limit as well as a BEC and the non-condensed fraction by more than half a millimeter. The transport and decompression of the atomic sample was engineered using dynamics invariants. Because of the Coulomb repulsion, the method used for cold atoms can not be extrapolated to ion clouds. For most experiments,  adiabatic transfer is not a relevant solution. For the trapping potentials in our experiment, the adiabatic transfer time is of the order of several tens of seconds, incompatible with a majority of precision experimental protocols. 

This article is organised as follows : the experimental set-up and techniques are presented in section \ref{s_setup}. Section \ref{s_transfer} is devoted to results and analysis of the transfer efficiency and the heating induced by the transport is analysed in section \ref{s_heating}. Section \ref{s_smallclouds} deals with the ion number effect for a specific transport duration with a focus on smaller clouds. Conclusion of this work can be found in section~\ref{s_conclusion}.

\section{Description of the experimental set-up and protocol} \label{s_setup}
\subsection{Trapping and laser cooling}
Calcium ions are trapped in a two-part linear quadrupole trap of inner radius $r_0=3.93$~mm which is designed to have reduced non-harmonic components in the trapping potential  : the RF-electrodes are four cylindrical rods of total length 58~mm 
 connected in a balanced way ($\pm (V_{RF}/2) \cos(\Omega t)$) to the RF supply (no grounded electrodes). The  RF-frequency, $\Omega/2\pi$,  is 5.23~MHz and the potential difference between neighbouring electrodes oscillates with a peak-to-peak amplitude of 1045~V (if not mentioned otherwise) which gives a Mathieu parameter of $q_x=0.15$, well within the adiabatic approximation regime where the RF-trapping can be approximated by a static harmonic potential (the pseudo-potential) with frequency $\omega_{x0}=q_x\Omega/(2\sqrt{2})=2\pi \times 277$~kHz.

Trapping along the symmetry axis is reached by DC-voltages applied to electrodes perpendicular to the rods. Three DC-electrodes are located at equal distances along the rods, creating two distinct trapping zones of 21~mm length, each. This double well configuration is used for accumulation in one of the wells before further transport of the ions to other traps, in line with the quadrupole  one. In \cite{kamsap15b} our protocol for genuine accumulation of ions is described in detail. In order to laser-cool the ions in both trapping regions using the same  laser beam, the DC-electrodes must leave the trap $z$-axis free, which justifies the open shape of the three of them. Their design, detailed in \cite{pedregosa10a}, results from a compromise between reduction of the non-harmonic contributions in the potential and of the screening effect  induced by the RF-rods.

Calcium ions are produced by photo-ionization of neutral calcium atoms from an effusive beam crossing  the trap axis perpendicularly in the horizontal plane. The photo-ionization process implies two photons and the first step (423~nm) is a resonant excitation tuned to select the most abundant isotope, $^{40}$Ca \cite{gulde01,lucas04}. The second photon, at 375~nm, takes the atomic system above the ionization threshold. Both beams co-propagate along the trap axis. Ions are laser-cooled by two collimated 397~nm-beams on the [$4S_{1/2} - 4P_{1/2}$]-transition, of equal power (2~mW  on a  2~mm $1/e^2$  diameter), counter-propagating along the trap axis. Once excited from the ground state, calcium ions can relax to a long-lived metastable  state [$3D_{3/2}$] from which they have to be re-pumped to maintain efficient laser cooling. This re-pumping process is assured by a 866~nm laser beam [$3D_{3/2} - 4P_{1/2}$] of approximately 2.5~mW and  4~mm $1/e^2$  diameter which co-propagates with one of the cooling lasers. Simultaneous ion creation and cooling allow  to trap clouds as large a several hundreds of thousands of ions. For the work presented here,  we tuned the ion creation parameters to reach a cloud size of the order of 20 000 ions, which takes typically an integration time of 15 seconds for laser powers of 80~$\mu$W at 423~nm and 4~mW at 375~nm. Ions are detected by their laser-induced fluorescence at 397~nm which is collected through a dedicated objective with anti-reflection coating  and a high numerical aperture (Sill Optics, $f=66.8$~mm, N.A = 0.28).

\begin{figure}
\begin{center}
\includegraphics[width=8.cm]{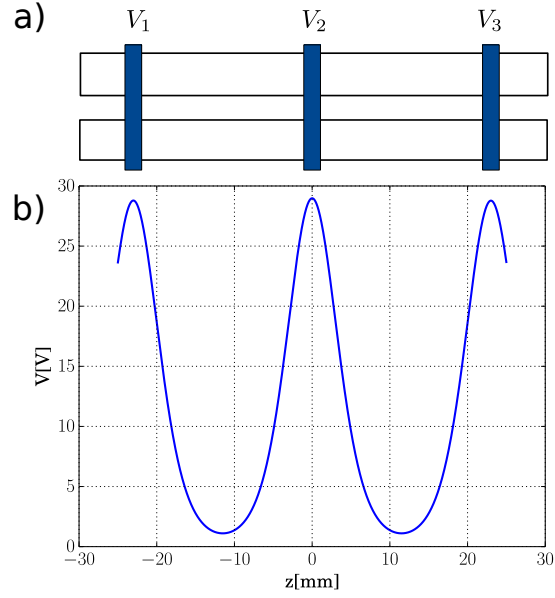}
\caption{(a): Scheme of the experimental set-up and (b): calculated DC- potential  along the trap axis if 1000V are applied to each electrode. The distance between electrode centers is 23~mm and the thickness of each electrode is 2~mm.}\label{fig_schemaVz} 
\end{center}
\end{figure}

The distance $L$ between the center of adjacent DC-electrodes is 23~mm and the trapping along the $z$-axis results from the electric potential gradient of the sum of each electrode contribution.  A solution of the Laplace equation by a finite difference  method software (Simion \cite{simion}) gives the potential profile associated to each electrode of thickness 2~mm. The characteristic shape of the resulting potential wells   along the trap $z$-axis are shown in Figure~\ref{fig_schemaVz}. The effective potential well, which can be estimated by the difference between the maximum and minimum total contribution, is lower than the DC voltage applied to the electrodes because of the screening by the RF-electrodes \cite{pedregosa10a} and by the overlapping of the potential profiles at the center of the trapping zones which offsets the potential minimum.  The large distance between electrodes results in a very small overlap of the profiles and as a consequence, in a low potential minimum. Nevertheless, the screening effect sets a limit  to the axial trapping efficiency : for 1000~V applied to each electrode, the voltage calculated at the electrode center is 29~V for the central one and 28.8~V for the end electrodes whereas the minimal potential value is 1.16~V. The small overlap which leads to a  deep potential well is a  drawback for transport issues, as explained in the following.

\subsection{Transport protocol}
As pointed out in the introduction, single ion transport without heating is a major issue for the scalability of trapped ion based quantum computer and is the subject of several  experiments \cite{reichle06,schulz06,hucul08}. In these works, the authors characterize and compare different transport protocols  with respect to  the  heating they induce on the ion motion. To guide us with our experiments, we have extrapolated these analysis for an ion cloud by using molecular dynamics (MD) simulations in \cite{pedregosa15}. We use the notations of previous work and  we call $\phi_i(x,y,z)$ the electric potential created by the DC-electrode $i$ when 1V is applied to it. Then, the total DC potential inside the trap can be expressed as \cite{singer10}
\begin{equation}
\Phi(t,x,y,z) = \sum_i^N{V_i(t)\phi_i(x,y,z)}
\end{equation}
if $V_i(t)$ is applied to electrode $i$. 

As the laser-cooled ions explore less than a tenth of the radial trap extension, we assume that the dependence of the DC-potential with the coordinates $x$ and $y$, perpendicular to the trap axis, is not relevant and we call $\phi_i(z)$ the on-axis evaluated function $\phi_i(x=0,y=0,z)$. With our electrode geometry,  the electric potential spatial distribution created on the axis  is very well fitted around its maximum by the equation~:
\begin{equation} \label{eq_fit}
f_i(z) = a_i\left(1+\frac{(z-z_{i})^2}{w_i^2}\right)^{-4}
\end{equation}
with $z_i$ the position of the center of electrode-$i$, $w_i=8.9$~mm and $a_{1,3}=28.8$~mV and  $a_{2}=29$~mV when $1~V$  is applied on the $i$-electrode.  The  non-symmetric environment of the trap explains the small variation between the $a_i$ values.  If $|z-z_i|=w_i$, $f_i(z)=a_i/16$ and we can consider that $2w_i$ is a good enough approximation of the effective width of the potential profile. When the three DC-electrodes are connected to the same potential, the potential well can be considered as harmonic around each potential minimum, behaving like $m\omega_z^2 (z-z_{c})^2/2$. The value of $\omega_z$ deduced from a fit of the potential around its minimum is $2\pi \times 107.5$~kHz for $V_i=1000$~V whereas a measurement by parametric excitation of the ion cloud \cite{champenois01} gives $2\pi \times 124$~kHz. The difference between these two values shows that the potential deduced from the calculation does not reproduce exactly the real potential experienced by the ions.

The transport protocol relies on the time variation of the $V_i$ potential,  designed to make the potential minimum obey a time profile $z_{min}(t)$. This condition translates into 
\begin{equation} 
\frac{\partial\Phi}{\partial z}\Big|_{z_{min}(t)} = 0 \label{eq_cond1}
\end{equation} 
There are two local minima which meet each other when they reach the centre of the central electrode and the challenge of the experiment is to design a potential evolution which transfers the ions from the minimum in zone 1 to the minimum in zone 2 (see figure~\ref{fig_schemaVz}). In the following, we call $z_{min}(t)$ the path we want the ions to follow and it can be written like \cite{hucul08}
\begin{equation}
z_{min}(t) = g(t)( H(t) - H(t-t_{g}) ) + L H(t-t_{g})-L/2  \label{eq_gate}
\end{equation}
with $L$ the shuttling distance, $g(t)$ the time profile of the transport, $t_g$ its duration, and $H(t)$ the Heaviside step function. Guided by numerical results detailed in \cite{pedregosa15},  we used for the experiments presented here the time profile described by
\begin{equation} 
g(t) =\frac{L}{2}\left(\frac{ \tanh\left( 4(2t/t_g - 1) \right)}{\tanh(4)} + 1\right). \label{eq_tanh}
\end{equation} 
These simulations showed that among  four compared  time profiles, the one following Eq.~\ref{eq_tanh} is the most robust against  transfer duration variations. They also give evidence that the deformation of the trapping potential along the transport is responsible for the heating of the center of mass motion and of the ions' motion in the center of mass frame \cite{pedregosa15} . In our experimental set-up, the distance between electrodes is larger than the effective width of the potential they create, therefore, keeping the axial potential undeformed while translating its minimum  requires huge voltages that we cannot provide. With the applied voltages  $V_i$ limited to 2000~V, we can only change the depth of the effective harmonic potential but cannot compensate for its deformation. 

In practice, the potential minimum is forced to obey the  time profile $z_{min}(t)$ if
 \begin{equation}\label{eq_V2}
V_2(t) = -\frac{V_1(t)\phi^{'}_1(z) + V_3(t)\phi^{'}_3(z)}{\phi^{'}_2(z)} \Big|_{z_{min}(t)}
\end{equation}
and the  harmonic contribution of the resulting axial potential along the transport can be deduced by 
\begin{equation} 
\omega_z^{2}(t)= \frac{Q }{m}\frac{\partial^{2}\Phi}{\partial z^{2}}\Big|_{z_{min}(t)}  \label{eq:cond2a}
\end{equation} 
where $Q$ is the ion charge and $m$ its mass.

Computing $V_2(t)$ through Eq.~\ref{eq_V2} requires to know the potential profile created by each electrode to estimate the first order derivative $\phi^{'}_i(z)$. Single-ion experiments have shown the great sensitivity of the transport induced heating on the precise knowledge of the potential geometry. Reference \cite{brownnutt12} proposes a characterisation method for micro-traps where a single ion explores the potential and its flaws. The size of our trap and of the ion sample are not suited for this method, and we have hence based our calculations on $\phi_i(z) \simeq f_i(z)$ (see Eq~\ref{eq_fit}).  Eq~(\ref{eq_V2}) leads to a discontinuity of $V_2(t)$ for $z_{min}(t)=z_2$, the center of the electrode 2 where $\phi'_2(z_2)= 0$. To avoid this discontinuity, a constant relation between $V_1(t)$ and $V_3(t)$ is imposed, given by
\begin{equation}
V_3 (t)= - V_1(t)\frac{f'_1(z_2)}{f'_3(z_2)} \label{eq:V3_1}
\end{equation}
In our experiments, $V_1$ and $V_3$ are kept constant. In a perfectly symmetric device $V_3(t) = V_1(t)$ would solve the problem but any asymmetry in the electrode environment breaks this equality. The reader is referred to \cite{pedregosa15} for details about how to avoid discontinuities in numerical simulations.

\section{Transfer efficiency for large ion clouds}\label{s_transfer}
\subsection{Estimation of the number of ions}
We are primarily interested in the transport efficiency as a matter of the relative number of ions passing from trapping zone 1 to zone 2. A precise quantitative study requires the measurement of the  number of trapped ions, which is of the order of a few tens of thousands.  The ion's fluorescence signal is split between a photomultiplier and an intensified CCD camera. For fixed laser frequencies and trapping parameters, the fluorescence counting rate depends on the number of ions in the trap but also on their temperature. Because the transport can induce heating and ion loss, and because RF-heating depends on the ion number and their temperature \cite{ryjkov05}, there is no simple relation between the recorded fluorescence signal and the number of trapped ions. Indeed, we very often observe a signal increase when the ion number has decreased, because of a smaller RF-heating. 

To develop a quantitative diagnostic independent of the signal counting rate, we use the density characteristics of the liquid phase of an ion cloud. One can show that in the cold fluid limit, a singly-charged sample in a harmonic potential has a uniform density \cite{prasad79}, bound by an ellipsoid of revolution where the density falls to zero on the scale of the Debye length. This results from the Boltzmann-Poisson equation in the low temperature limit and for ions in a linear quadrupole trap the density depends only on the trapping pseudo-potential   \cite{champenois09}. This property, as well as the predicted aspect ratio of the ellipsoid \cite{turner87},  have been verified quantitatively very accurately for ions in a linear quadrupole trap in \cite{hornekaer02}.

Every ion ensemble is cooled  to the liquid phase before and after transport in order to quantify the ion number. The difference between liquid and gas phase is easily detected by the variation in the fluorescence level \cite{hornekaer02}. The calculated density in the liquid phase is $1.40\times10^{5}$ mm$^{-3}$ and for a typical temperature of 100~mK, the Debye length is $1.85~\mu$m, which fits within two pixels on the camera with  an optical magnification of the order of 13. The typical size for the ion cloud in the liquid phase is of the order of 700~$\mu$m for the  semi-major axis and 300~$\mu$m for the  semi-minor axis of the ellipse. The Debye length is then negligible compared to the cloud size and we consider a uniform density all over the cloud in the liquid phase. 
 By changing the laser cooling efficiency and thereby the sample temperature, we checked that, for ion clouds smaller than 50 000 ions,  once the ellipsoid is formed by laser cooling, the measured dimensions for the ellipse  are independent of the fluorescence signal level and the slight modification of the Debye length has no impact on the measured values.  For larger clouds, transition to the liquid phase requires to reduce the RF-voltage amplitude and the border of the ellipses spread over a larger scale. We have developed a software which  automatically fits and extracts the dimensions  of an ellipse from the recorded picture of the collected fluorescence and the method used by this software is now developed.

\begin{figure}[htb]
\begin{center}
\includegraphics[width=5.cm]{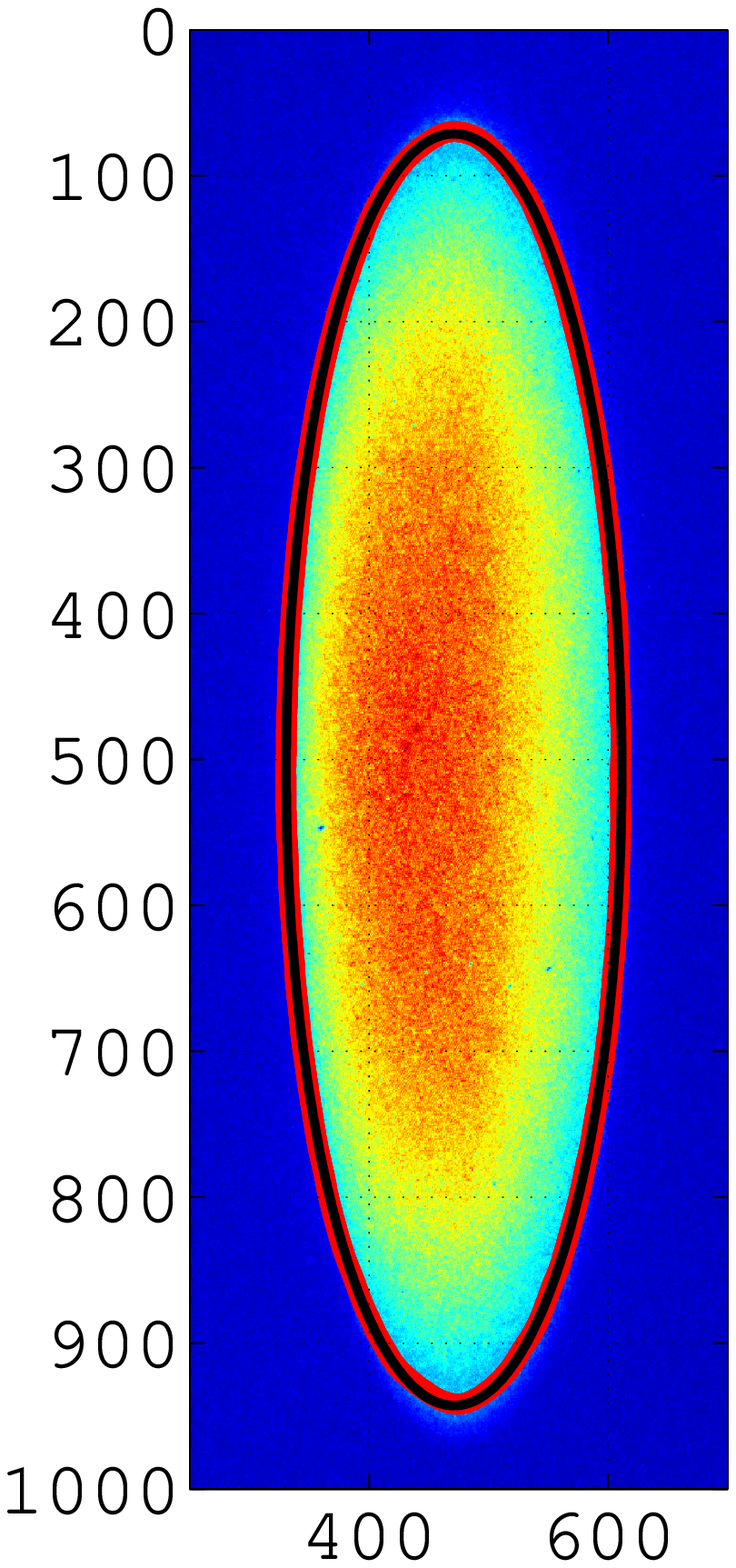} \includegraphics[width=8.cm]{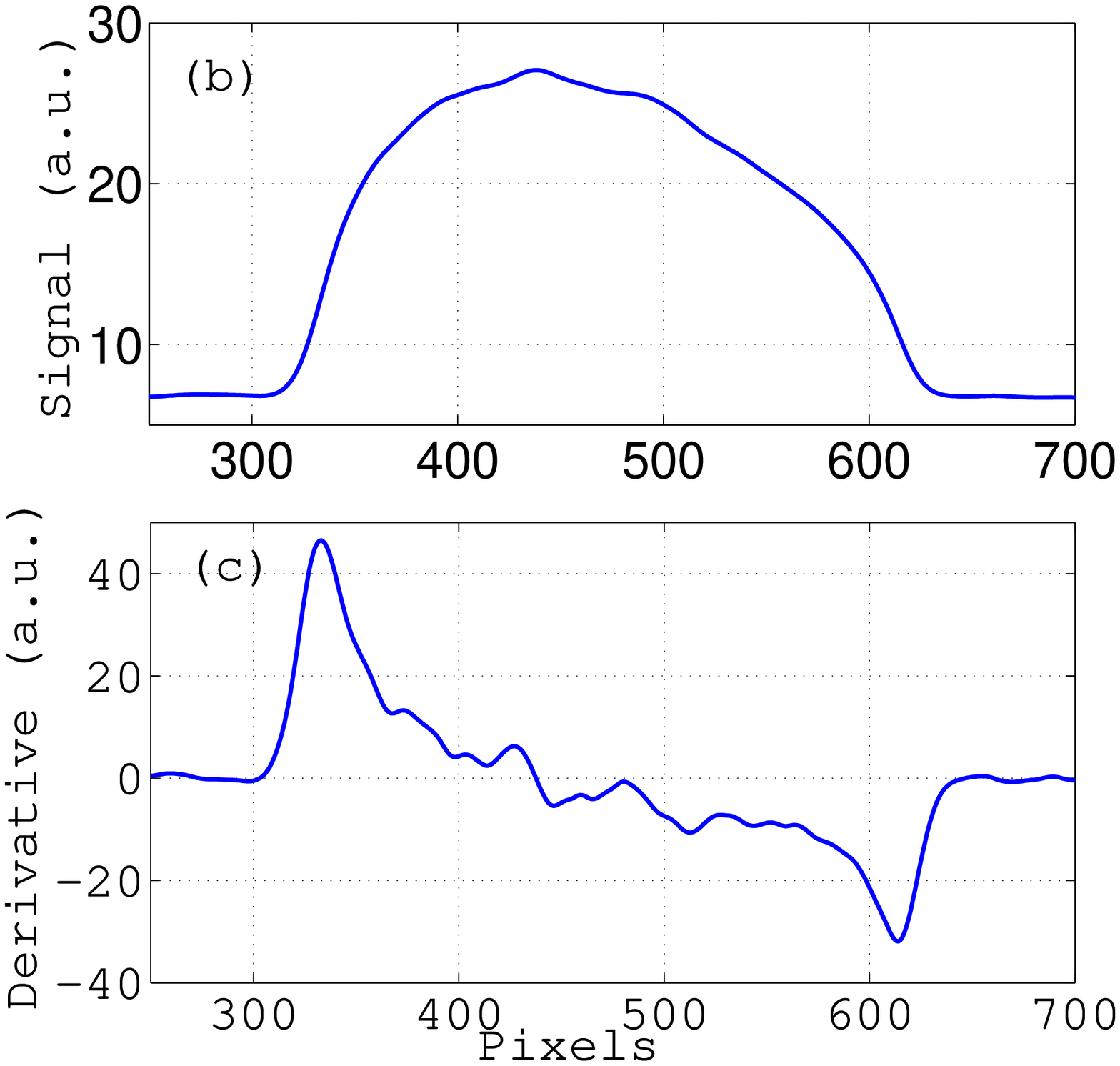}
\caption{(a) example of the picture of an ion cloud (dimensions in pixels) in the liquid phase with the fit of the contour (exposure time 0.5~s). The trap axis and laser propagation direction are vertical. (b) Smoothed section of the signal along the short axis of the ellipse, and (c) its derivative }\label{fig_threshold}
\end{center}
\end{figure}

 The first step is to define the contour of the fluorescence signal. This requires the definition of a threshold for the signal, independent from the number of photons scattered per ion. The threshold criteria is provided by the analysis of the section  of the signal along one pixel line across the image. The derivative of the signal with respect to the pixel position shows two sharp extrema, $X_1$ and $X_2$  at the ellipse border (see figure \ref{fig_threshold}). Their position falls in the same pixel as the one chosen by a fit "by the eyes" and does not depend on the  absolute level of the signal. As some of the ellipse pictures are longer than the camera detector, we base our protocol on a section located approximately on the small axis and define the level threshold as the mean signal $(S(X_1)+S(X_2))/2$.

 The second step is to fit the contour by an ellipse equation, including a possible angle between the ellipse semi-axis and the detector rows. This is done through a 2D fit subroutine  and produces a fit which falls in the same pixels as the original contour, which confirms the ellipsoid shape of the cloud. To valid our fit procedure, for the same ion cloud, simply deformed by changing the trapping parameters, we use the two calculated semi-axis lengths $R_e$ and $L_e$ to estimate the ellipse aspect ratio $\rho_e=R_e/L_e$ and  volume $V_e=4 \pi R_e^2L_e/3$ extracted from experimental data. We can compare the aspect ratio with the expected one, deduced from the effective pseudo-potential $\rho=f(\omega_z/\omega_r)$ by an equation demonstrated in \cite{turner87} and experimentally confirmed in \cite{hornekaer01}. More precisely, we measure the relative shift $\delta_L$ between the  length deduced from the fit and the one deduced by the fitted radius, assuming a known aspect ratio : $\delta_L=(L_e-R_e/\rho)/L_e$.  Furthermore,  we can check that the volume modifications obey what is expected from the density $n({\mathbf r})$ at low temperature :
 \begin{equation}
n({\mathbf r})=\epsilon_0 \Delta \Psi_{pp}({\mathbf r})/Q
\end{equation}
where  $\Psi_{pp}({\mathbf r})$ is the harmonic pseudo-potential, characterised by 
 \begin{equation}
 \Psi_{pp}({\mathbf r})=\frac{1}{2Q} m(\omega_x^2-\omega_z^2/2)(x^2+y^2)+\frac{1}{2Q} m \omega_z^2 z^2
\end{equation}
which leads to a uniform density $n({\mathbf r})=n_0=\epsilon_0 2m\omega_x^2/Q^2$. As we want to reach a 1\% level accuracy in relative volume estimation, we have to go beyond the first order adiabatic approximation. By expanding the calculation of the coefficient in the Mathieu solutions to the second order in ($q_x^2, a_x$), one can show that
\begin{equation}
\omega_x^2=\omega_{x0}^2\left(1+\frac{q_x^2}{2}+a_x\right).
\end{equation}
In our case, $a_x$ is induced by the $z$-axis trapping-voltage $V_{DC}$, $a_x=-2\omega_z^2/\Omega^2$ and with our operating parameters, the correction is in the 1\% range. Taking that into account, for the same RF amplitude but different DC-voltages, we observe  relative fluctuations  lower than $\pm 1$\% for both the length and the volume of the same cloud in zone 2. In zone 1, we observe volume fluctuations that can reach 6 \%, far larger than the length fluctuations $\delta_L$ which remain in the $\pm 1$\% range. We attribute this difference to the contact potential, identified in zone 1, and induced by calcium deposition on the quadrupole rods, in front of the calcium oven. For different trapping parameters, the cloud is displaced in the trap, giving an optical image with a slightly different size. This is in particular true when the RF amplitude is changed and where apparent ion numbers can vary by 10\%. For constant trapping parameters, like used for estimating the relative number of ions after a transport protocol, the uncertainty on the volume is $\pm 1$\% for zone 2 and  $\pm 1.5 $\% for zone 1 ($6\sigma$ confidence). Precise investigations of transport efficiencies in terms of ratios of ion numbers, only requires measurement of the volume of the ellipses as we compare cold clouds with identical density. As for an estimation of the  number of ions in a cloud, the fluctuations of the apparent particle number for the same cloud, when it is deformed and shifted, lead us to fix a 5\% uncertainty on the absolute number (and the uncertainty induced by the optical magnification is negligible here).

\subsection{From one local minimum to the other}
In a first step, we study our ability to transfer ions from one trapping zone to the other, depending on the duration of the transfer $t_g$ for  given trapping parameters.   Bandpass limitations of the DC-supplies prevent us from investigating  transport durations shorter than 80~$\mu$s. All the experiments were done with an ion cloud with a typical size ranging from 5000 to 20000 ions. For some well identified transport protocols, we checked that the transport efficiency is independent of the ion number as long as this number is larger than 2000. For clouds smaller than 2000, the efficiency is higher than for larger clouds and can reach 100~\%. A focus on smaller cloud transfer is presented in section \ref{s_smallclouds}. As our detection is based on the observation of induced fluorescence, the  cooling laser beams  remain applied during the transport. For some specific transport protocols, we compared the transport efficiency with and without the cooling laser during transport. The observed differences were only of the order of a  few~\% showing that the cooling effect does not play an important role.  Indeed, the capture range of the Doppler laser cooling is 9.2~m/s, smaller or far smaller than  the average shuttling velocity which ranges from 20 to 200 m/s. 

\begin{figure}[htb]
\begin{center}
\includegraphics[width=15.cm]{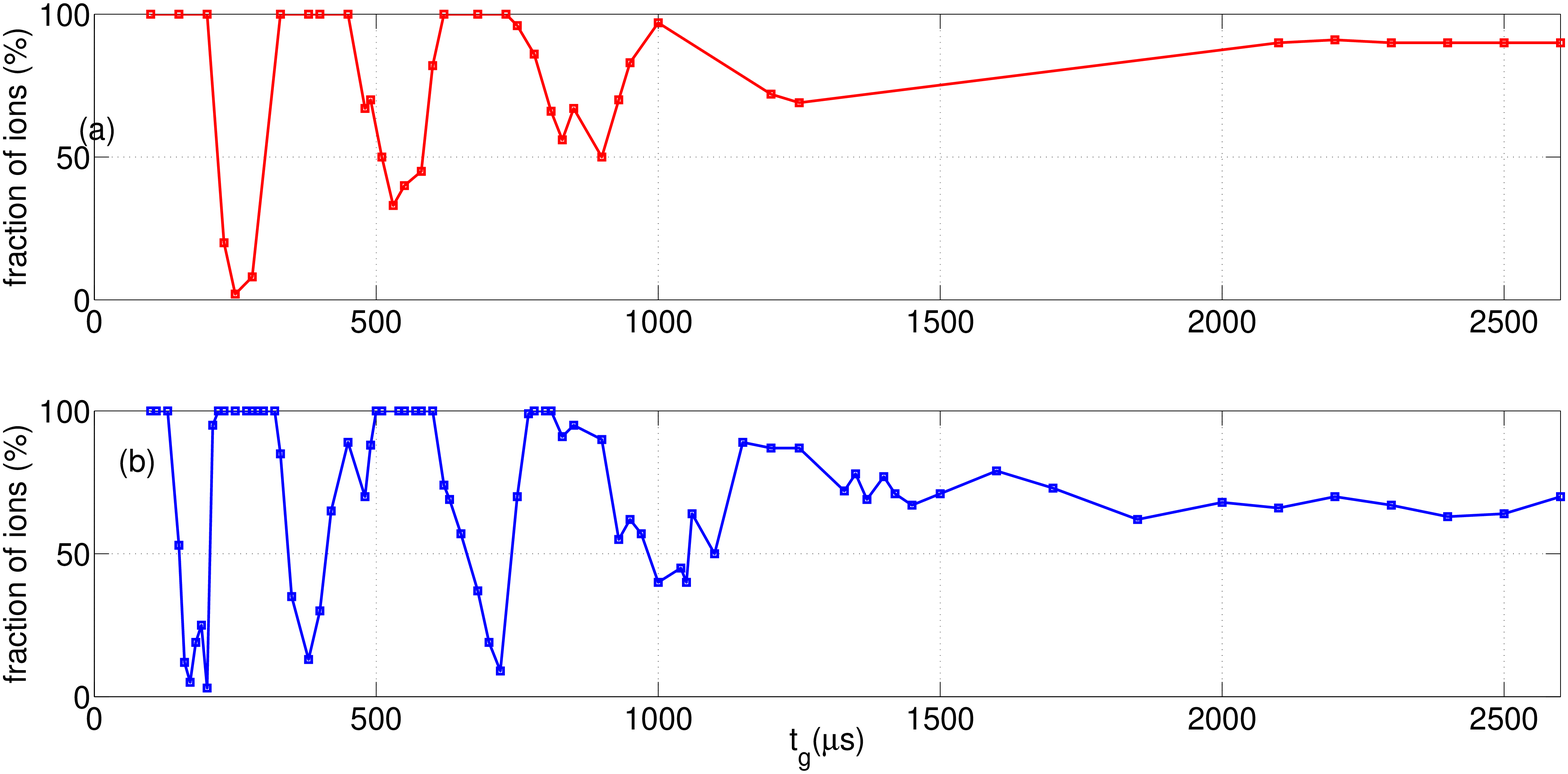}
\caption{Fraction of ions leaving the  trapping zone 1 vs the transport duration $t_g$ for different on-axis DC-voltages  : (a) $V_i=600$~V, (b) $V_i=1200$~V. The lines are a means to guide the eye. }\label{fig_depart}
\end{center}
\end{figure}

Figure~\ref{fig_depart} shows the fraction of ions leaving the trapping zone~1 for  $t_g$  between 80$\mu$s and 2.6~ms. The first major observation is that the number of leaving ions depends strongly on $t_g$, alternating between nearly 0 and 100\% several times before these oscillations are damped. 
Changing the axial trapping potential by changing the DC voltages, shifts these oscillations with $t_g$  and makes more oscillations observable for a larger value of the DC voltages. If the same protocol is applied to a smaller ion cloud (typically 1000 ions and smaller), oscillations are also observed with identical temporal imprint, excluding a number dependent effect. A possible explanation for this interchange between a high and low transfer probability    is the   oscillation of the ions  from zone 1 to zone 2 and their return to trapping zone 1 before the transport function is completed. This assumption is tested by setting $V_3$ to -2000~V during the transport protocol, which means opening the second trapping zone. For this configuration, oscillations are still observed, shifted in $t_g$ with respect to the previous symmetric configuration. These observations are consistent with the hypothesis that ions that were still in trapping zone~1 once the transport protocol was completed, entered   zone~2 far enough to see their trajectories modified by the deformation of the potential, but not far enough to be attracted by the negatively polarised electrode. For transport durations longer than 1~ms, the fraction of transferred ions reaches a stationary value. In an ideal, symmetric system this value is expected to be 50 \%. In our dual trap,  an asymmetry, very probably due to the  contact potential in trapping zone 1, can be responsible for this imbalance in ion repartition for long transport.

\begin{figure}[htb]
\begin{center}
\includegraphics[width=13.cm]{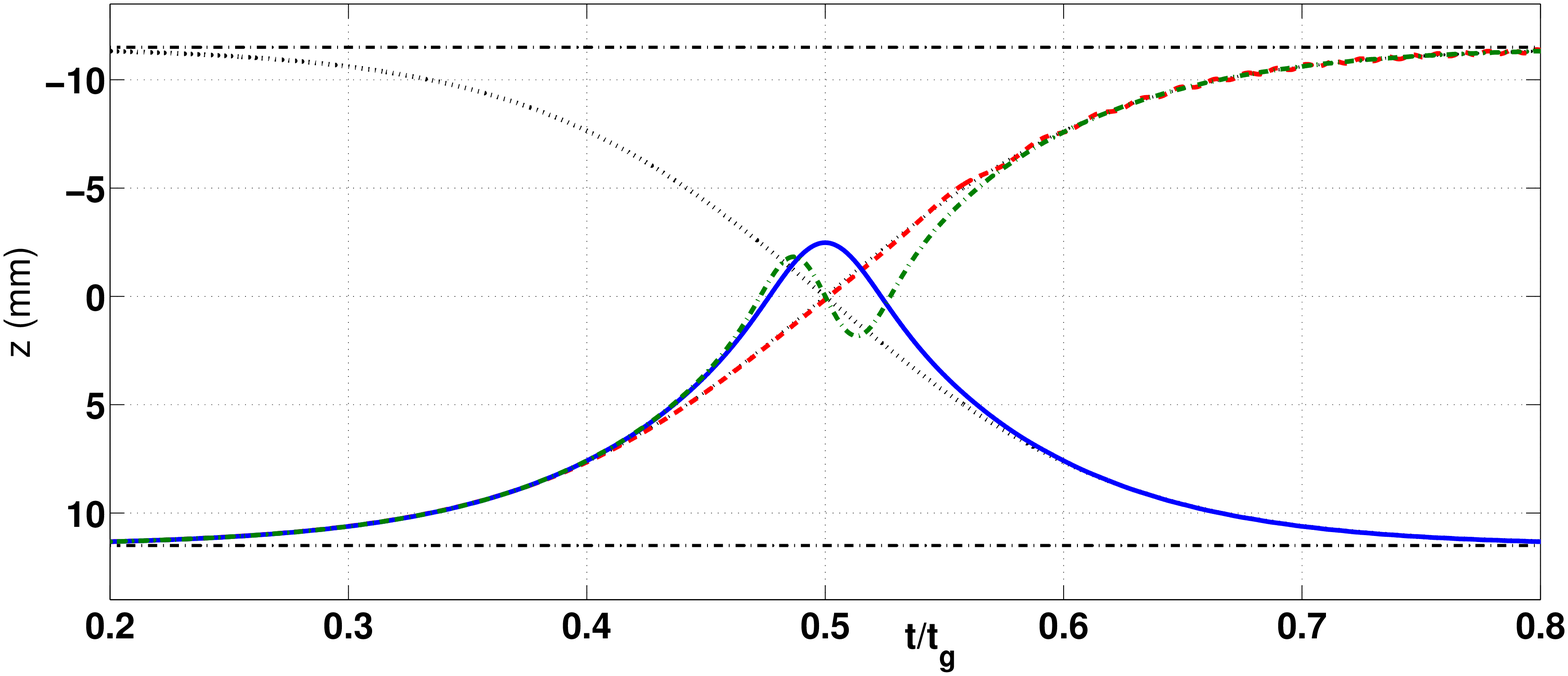}
\caption{ Single ion trajectories  computed by MD, versus the  relative time elapsed during the  transfer. The  trajectories are computed in the potential grid provided by the Simion software, based on our electrode geometry, when $V_2(t)$ obeys Eq.~\ref{eq_V2} computed with $\phi_i(z)=f_i(z)$. Red dashed line  : $t_g=189~\mu$s, solid blue line :  $t_g=884~\mu$s,  green dot-dashed line : $t_g=1621~\mu$s. The black dotted lines are the position of the two potential minima along the transfer. Horizontal dash-dotted lines indicate the positions of the centres of the traps. }\label{fig_traj}
\end{center}
\end{figure}

To get more insight into this issue, we use a MD simulation to compute the trajectory of a single ion in the translated and deformed potential applied in the experiment, as a good approximation of the center of mass motion \cite{pedregosa15}. Actually, the results of this simulation depend on the method used to describe the potential. If the equations of motion are integrated  in the calculated axial potential fitted by Eq.~\ref{eq_fit}, the probability for the ion to be transferred to zone 2 is unity, whatever is the transfer duration.  To come closer to the experimental situation, we keep the waveforms $V_i(t)$ as used in the experiments but integrate the equation of motion in the potential grid calculated by the finite difference method software, Simion \cite{simion}. In this condition, oscillations of the probability to transfer the ion  to trapping zone~2 are observed, depending on the transfer duration. The consequence of the discrepancy between the two descriptions of the potential is visible on figure~\ref{fig_traj} where several examples of ion trajectories are ploted.
For $t_g=189~\mu$s, the ion trajectory follows the potential minimum from trapping zone~1 to zone~2. For longer transfer durations, the ion is ahead of the potential minimum and  for $t_g=884~\mu$s, makes a U-turn in zone~2 before ending in zone~1, like assumed previously. For even longer transfer times ($t_g=1621~\mu$s), the calculated trajectory shows a U-turn in zone~2, followed by a U-turn in zone~1  to finally have an ion efficiently transferred to zone~2. Longer transfer durations give rise to an increasing number of U-turns which results  in an oscillation between trap~1 and trap~2 for the ion final position. The experimental results exhibit  a shorter timescale than the simulations. This can be explained by a larger difference between the potential assumed to compute the $V_i(t)$ and the one experienced by the ions in the trap. The inconsistency identified above cannot be avoided in a large-scale experiment where it seems unrealistic to generate a precise map of the complete electric field seen by the ion cloud along its transport. Nevertheless, the experimental results show that it is still possible to force the ion to transfer even if the corresponding  time scale can not be exactly foreseen. The next step in the transport efficiency analysis is to look at how many ions effectively settle in the other part of the trap.

\subsection{Transport induced ion loss }
\begin{figure}[htb]
\begin{center}
\includegraphics[width=12.cm]{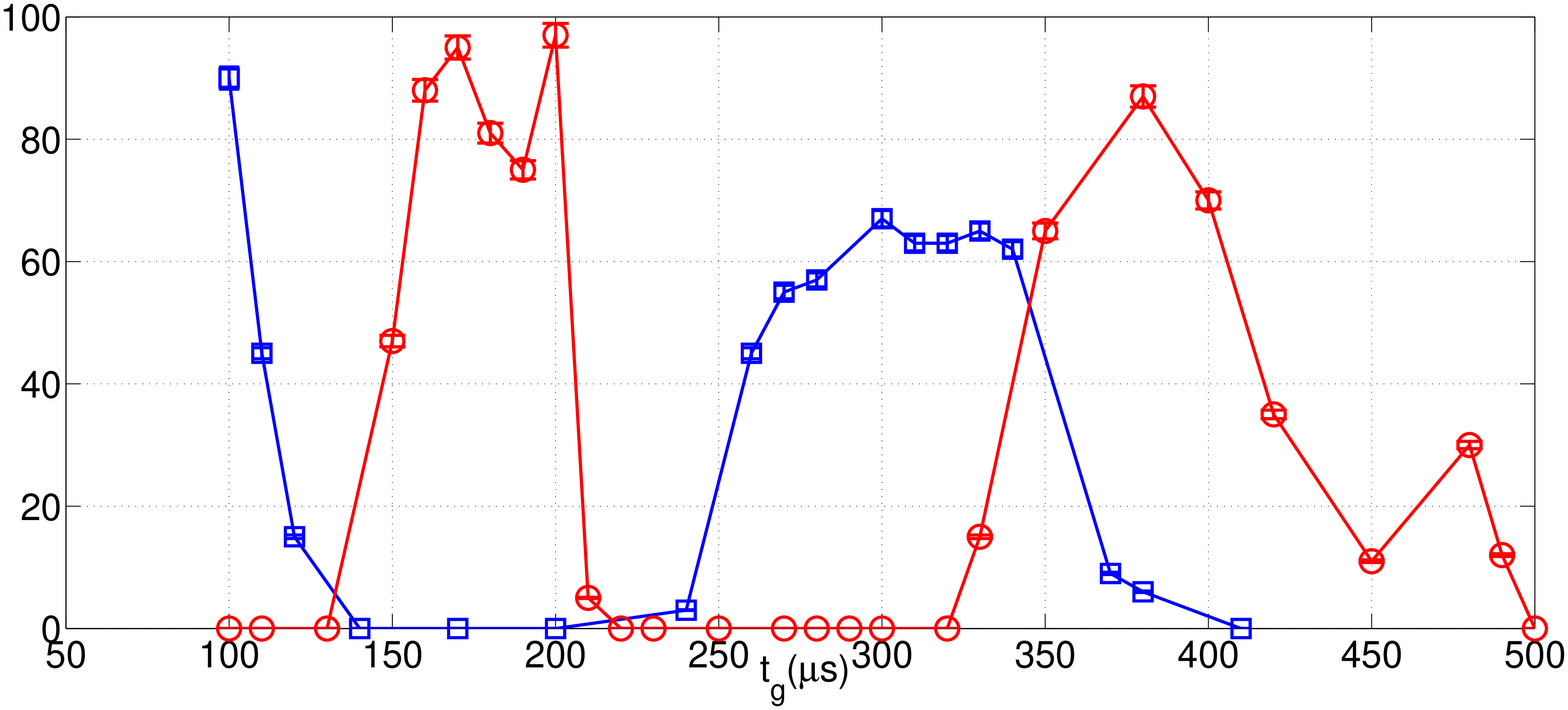}
\caption{Blue squares : Proportion of ions transferred from zone~2 to  zone~1 by a transfer protocol of duration 100~$\mu$s  and then transferred back to zone~2 by a transfer protocol of duration $t_g$. Red dots (deduced from Fig.~\ref{fig_depart}(b)): Proportion of ions not leaving zone~1 after transfer protocol of duration $t_g$. The DC voltages are 1200~V (the lines are a guide to the eye) }\label{fig_transfer2to1}
\end{center}
\end{figure}

The third  step in our transport efficiency characterisation is to check that all ions leaving zone~1 are trapped in zone~2 by the end of the transport protocol. As only a single fluorescence collecting optics is used in the experiment, the precise characterisation of the transfer efficiency requires that the ions are transferred back to their original position for a comparison between the cloud sizes.  To circumvent this drawback and be able to estimate the one-way transfer efficiency, we identified a transfer protocol that is efficient  enough to serve as a standard operation. This is the case for the   transfer of 100~$\mu$s duration. Like mentioned previously, the transfer efficiency depends very little on the ion number as long as this number is larger than two thousand ions. This efficiency was estimated from several consecutive transport protocols to be of the order of 90\% for 100~$\mu$s. By using always the same protocol for the zone~2 to zone~1 transfer, we can observe the dependence   of the zone~1 to zone~2 transfer efficiency as a function of its duration,  like shown on figure \ref{fig_transfer2to1}.
The results show oscillations with the transfer duration, out of phase with the proportion of ions not leaving trapping zone~1, which is also reproduced from Fig.\ref{fig_depart}(b) on the same figure. The largest two-way transfer efficiency is as high as 90\% and is observed for a transport made of two consecutive 100~$\mu$s transport protocols. Increasing the duration of a transport protocol does not result in a higher transfer efficiency. Numerical simulations detailed in \cite{pedregosa15} show how the cloud spreading makes  long transport inappropriate for large clouds. The extra information brought by the comparison of the two curves of Figure~\ref{fig_transfer2to1} is that the ion number budget evidences transfer-induced ion loss. In the following we quantify the  transfer-induced  cloud heating to look for possible  correlations with the ion loss.


\section{Excitation of motion}\label{s_heating}

\begin{figure}[hb!]
\begin{center}
\includegraphics[width=14.cm]{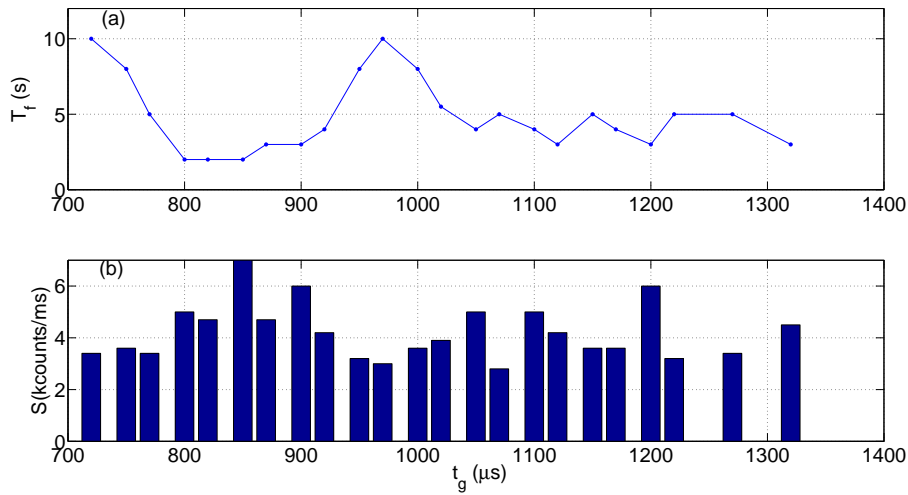}
\caption{a): Time $T_f$ it takes for a transported cloud in zone~2 to recover its maximal fluorescence rate after a transfer from zone~1 of duration $t_g$.  b): Value of this maximal signal. The DC initial voltages are 1800~V. }\label{fig_cooling_time}
\end{center}
\end{figure}

The transport-induced motional excitation has a signature on the time evolution of the  fluorescence. All laser frequencies are kept constant during the experiment and the temporal evolution of the  fluorescence  directly after the transport depends on the Doppler effect, which depends on the ion velocity along the trap axis.  Heating may occur during transport, in that case, a re-cooling phase can be observed. The time $T_f$ required for the fluorescence rate to reach its stationary value after a transfer operation is plotted in figure~\ref{fig_cooling_time}. The chosen transport durations are large enough to make sure that a non-negligible proportion of the ion cloud arrives in zone~2.  The re-cooling time $T_f$ can vary from  short (2~s) to long (10~s) times which shows that the Doppler shift induced by the transport depends on the duration of the transport. Also on figure~\ref{fig_cooling_time} (b) is plotted
the maximum fluorescence rate, which can be considered as a crude indication for the ion number. The graph confirms that a fast recovery of the signal is not due to a lower number of ions. As seen on the comparison of the two curves of  figure~\ref{fig_cooling_time}, the amplitude variations of the signal are anti-correlated with the recovery time of the fluorescence rate. In a hand-waving argument, we can interpret the time $T_f$ as an indicator for  the motional excitation, and deduce from this figure  that a larger number of ions is efficiently transferred to trapping zone~2 when this excitation is low. 
 When a smaller RF amplitude is used (which results in $q_x=0.12$ instead of 0.15), the signal recovery time can be as small as 200~ms, showing that the ion velocities are less modified by the transport process. The largest observed signal recovery time is  5~s for this lower potential value. The global increase of the signal recovery time with the RF-amplitude can be interpreted by non-linear terms in the equation of motion coupling the motion along the radial and axial directions, and giving rise to RF heating of the  motion.  In our experimental context, the axial potential is deformed and the non-harmonic contributions are non-negligible when the potential minimum crosses the site of the central electrode. The anharmonic contributions induce a coupling between the center of mass motion and the motion in the center of mass frame which  is responsible for an increase of the kinetic energy of the center of mass. MD simulations of the transport of an ion cloud \cite{pedregosa15} showed that this contribution increases with the transport duration, as the cloud spreads further out. The experimental results do not show such a behaviour for the time-scale explored, leaving the cause of duration-dependent cloud heating unexplained.

\section{Transfer efficiency versus ion number}\label{s_smallclouds}
\begin{figure}[htb]
\begin{center}
\includegraphics[width=14.cm]{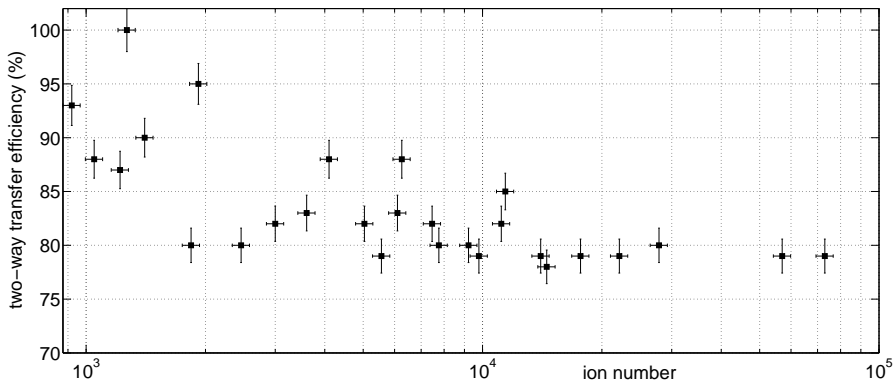}
\caption{Ratio of the number of ions after shuttling back and forth from trapping zone 2 with two identical transport protocols, versus the initial number of ions.\label{fig_smallclouds}}
\end{center}
\end{figure}

We analyse the ion number effect for the transport duration which gives the highest two-way transport efficiency. In the described case, the protocol uses transport functions of duration $t_g=100 \mu s$. Figure \ref{fig_smallclouds} shows that for  clouds of less than 5 000 ions, the round-trip transfer efficiency increases with shrinking cloud size and can reach  unity for ensembles of less than 2000 ions. We  assume that this size effect is due to the spatial  spreading of the cloud. This figure also shows that for this chosen transport function, the round-trip efficiency for shuttling is typically higher than 80  \%.  These high ratios can be realised with ion clouds of up to 10$^5$ ions. This very fast and efficient shuttling is in particular interesting for experiments in frequency metrology, as for example \cite{prestage07}.

\section{Conclusion}\label{s_conclusion}
We have experimentally studied the transport of ion clouds in a macroscopic rf trap for cloud sizes as large as several tens of thousands.  This transport is efficiently controlled by the potential applied to the central DC-electrode splitting the trap in two zones. We  have used a  time profile for the transport function  which is designed for single ion shuttling, and which can result in unity transfer of an ion cloud. We have also investigated the cloud's response to the potential deformation induced by the  spatial translation. Our experimental results and their comparison with MD simulations show that the mismatch between the ideal and the real DC-potential profiles explains the varying temporal response of the transport efficiency on the duration  of the transport function.  We can observe oscillations between high and low probability for an ion-cloud transfer from one trapping zone to another. 
The transfer induced heating also shows such oscillations with a comparable time scale.

The observed oscillations in the transfer efficiency can be modified by choosing different trapping parameters, and the minimum heating can be lowered by using smaller RF amplitudes. It is therefore  possible to find conditions for which the transfer efficiency is high and the motional excitation is low for the same transfer duration.  For clouds containing less than 2000 ions, 100 ($\pm 1.5$)\%-transfers can be achieved. This is another step approaching our objective which is to transfer large ion clouds with 100\% efficiency without heating. Our best results for clouds  larger than 5000 ions, are transfers of 92\%. 

We can tailor the temporal response to the transfer protocol by tuning the applied DC-voltages. It is possible to choose a "no-return" parameter set, where ions are transferred  with a very high probability from a first trap to a second trap, but at the same time they do have an extremely low probability to leave the second trap. This asymmetric protocol allows to implement a true accumulation process, the experiment is described in \cite{kamsap15b}.

\begin{acknowledgments}
We acknowledge the technical support of \'{E}meline Bizri and Vincent Long during the design and construction of the set-up, and of Stahl Electronics for the development of the dedicated RF sources.
This experiment has been financially supported by ANR (ANR-08-JCJC-0053-01), CNES (contract n$^{\circ}$116279) and R\'{e}gion PACA. MRK  acknowledges financial support from CNES and R\'{e}gion Provence-Alpes-Cote d'Azur.
\end{acknowledgments}

%

\end{document}